\documentclass[10pt]{article}
\usepackage{amsmath}
\usepackage{graphicx}
\usepackage{verbatim}
\usepackage{comment}

\begin{document}


\title{Measuring Form Factors and Structure Functions with CLAS}


\author{G.P. Gilfoyle  (for the CLAS Collaboration)}





\begin{center}

{\bf \Large Measuring Form Factors and Structure Functions \\with CLAS}

\end{center}

\vspace{0.25in}

\noindent G.P. Gilfoyle

\noindent representing the CLAS Collaboration

\noindent Physics Department, University of Ricmond

\noindent 28 Westhampton Way

\noindent Richmond, VA 23173 USA

\bigskip

\noindent Proceedings of the third International Conference in High-Energy Physics: HEP MAD07, \\
27 Sep - 2 Oct 2007, Antananarivo, Madagascar.

\begin{center}
{\bf Abstract}
\end{center}

\begin{changemargin}{0.25in}{0.25in}
The physics program at the Thomas Jefferson National Accelerator Facility
includes a strong effort to measure form factors and structure functions to
probe the structure of hadronic matter, reveal the nature of confinement, and develop
an understanding of atomic nuclei using quark-gluon degrees of freedom.
The CLAS  detector is a large acceptance device occupying one of the end stations.
We discuss here two programs that use CLAS; measuring
the magnetic form factor of the neutron, and the virtual photon asymmetry
of the proton.
The form factor has  been measured with unprecedented 
kinematic coverage and precision up to $\rm Q^2=4.7~GeV^2$ and is
consistent within 5\%-10\% of the dipole parameterization.
The proton virtual photon asymmetry has been measured across a wide range
in Bjorken $x$.
The data exceed the $SU(6)$-symmetric quark prediction and show evidence of a
smooth approach to the scaling limit prescribed by perturbative QCD.

\end{changemargin}


\vspace{0.25in}

\section{Introduction}

The Thomas Jefferson National Accelerator Facility (Jefferson Lab or JLab) is the United States' newest
national laboratory and is located in Newport News, VA.
It is focused on mapping the geography of the transition from the successful hadronic 
model of atomic nuclei to one based on the underlying quark-gluon constituents of matter.
The central instrument is the Continuous Electron Beam Accelerator Facility (CEBAF) which
is a superconducting,  linear,  electron accelerator.
About 1.4-km long and shaped like a racetrack, it can produce electron beams up to 6 GeV in
energy with 80\% polarization.
Currents can vary from 1-50 nA.
One of the accelerator end stations houses the CEBAF Large Acceptance Spectrometer (CLAS), a 35-ton, 
large-solid-angle device built around six superconducting coils that produce a toroidal
magnetic field \cite{nim}.
The CLAS consists of layers of drift chambers to measure charged particle trajectories, 
scintillators for timing measurements, Cerenkov counters to identify electrons, and electromagnetic
calorimeters to measure energy.
The focus of this paper is the measurement of form factors and structure functions with CLAS.
Below we focus on two recent experiments in CLAS to measure the neutron magnetic form factor and
the proton virtual photon asymmetry.

\section{Magnetic Form Factor of the Neutron}

The elastic form factors of the proton and neutron are fundamental
quantities which have been studied for decades. The dominant
features of the larger form factors $G_M^p$, $G_E^p$, and $G_M^n$ were
established in the 1960's: the dipole form $G_{D} = 
(1+Q^2/0.71)^{-2}$ gave a  
good description within the experimental uncertainties, corresponding 
(at least for $Q^2<<1 ~GeV^2$) to an exponential falloff in the spatial
densities of charge and magnetization. In the intervening decades,
obtaining higher precision measurements of these quantities has been
one thrust of the field, while new directions have also emerged,
especially over the past decade. These include precise measurements of
the neutron electric form factor~\cite{Gen}, and
extractions of the strange 
electric and magnetic form factors for the proton~\cite{G0}, as well as
time-like form factors~\cite{timelike}. In
addition to experimental progress, there has been renewed theoretical
interest on several fronts~\cite{Kees}. First, models of the nucleon
ground state can often be used to predict several of these quantities,
and it has proven to be very difficult to describe all of the modern data
simultaneously in a single model approach. Second, lattice
calculations are now becoming feasible in the few-GeV$^2$ range, and
over the next decade these calculations will become increasingly
precise. Finally, since elastic form factors are a limiting case of
the generalized parton distributions (GPDs), they can be used to
constrain GPD models \cite{Kroll}. For this purpose, high precision and a large
$Q^2$ coverage is important~\cite{Kroll}. At present the neutron
magnetic form factor at larger $Q^2$ is known much more poorly than
the proton form factors. 


The present measurement~\cite{jl1} makes use of quasielastic scattering on
deuterium where final state protons and neutrons are detected. The
ratio of $^2\rm{H}(e,e'n)$ to $^2\rm{H}(e,e'p)$ in quasi-free kinematics is
approximately equal to the ratio of elastic scattering from the free
neutron and proton. The ratio is: 

\vspace{-0.6cm}

%
\begin{eqnarray}
R_D ~~ = ~~ 
{{d\sigma \over d\Omega}[^2\rm{H}(e,e'n)_{QE}] \over 
{d\sigma \over d\Omega}[^2\rm{H}(e,e'p)_{QE}]
}
~~ = ~~ a \cdot R_{free} ~~ = ~~ a \cdot { 
{(G_E^n)^2+\tau(G_M^n)^2 \over 1+\tau}
+2\tau(G_M^n)^2\tan^2({\theta\over2})
\over 
{(G_E^p)^2+\tau(G_M^p)^2 \over 1+\tau}
+2\tau(G_M^p)^2\tan^2({\theta\over2})
}
\end{eqnarray}
%
\vspace{-.4cm}

\noindent where $\tau = Q^2/4M^2$, $M$ is the nucleon mass, and $\theta$ is the electron scattering angle.
Using deuteron models one can accurately compute the correction
factor $a(Q^2,\theta _{pq})$,
which is nearly unity for quasielastic kinematics and higher
$Q^2$. The value of $G_M^n$ is then obtained from the measured value
of $R_D$ 
and the experimentally known values of $G_E^n$, $G_M^p$, and
$G_E^p$.
This method has been used previously~\cite{Sick}. 
The $(e,e'n)$ and
$(e,e'p)$ reactions were measured in this work at the same time from the same
target. Use of the ratio $R_D$ under these circumstances
reduces or eliminates several experimental uncertainties,
such as those associated with the luminosity measurement or radiative
corrections. The remaining major correction is for the detection 
efficiency of the neutron. 

\begin{figure}[htb]
\hspace{0.5in}\includegraphics[height=1.8in]{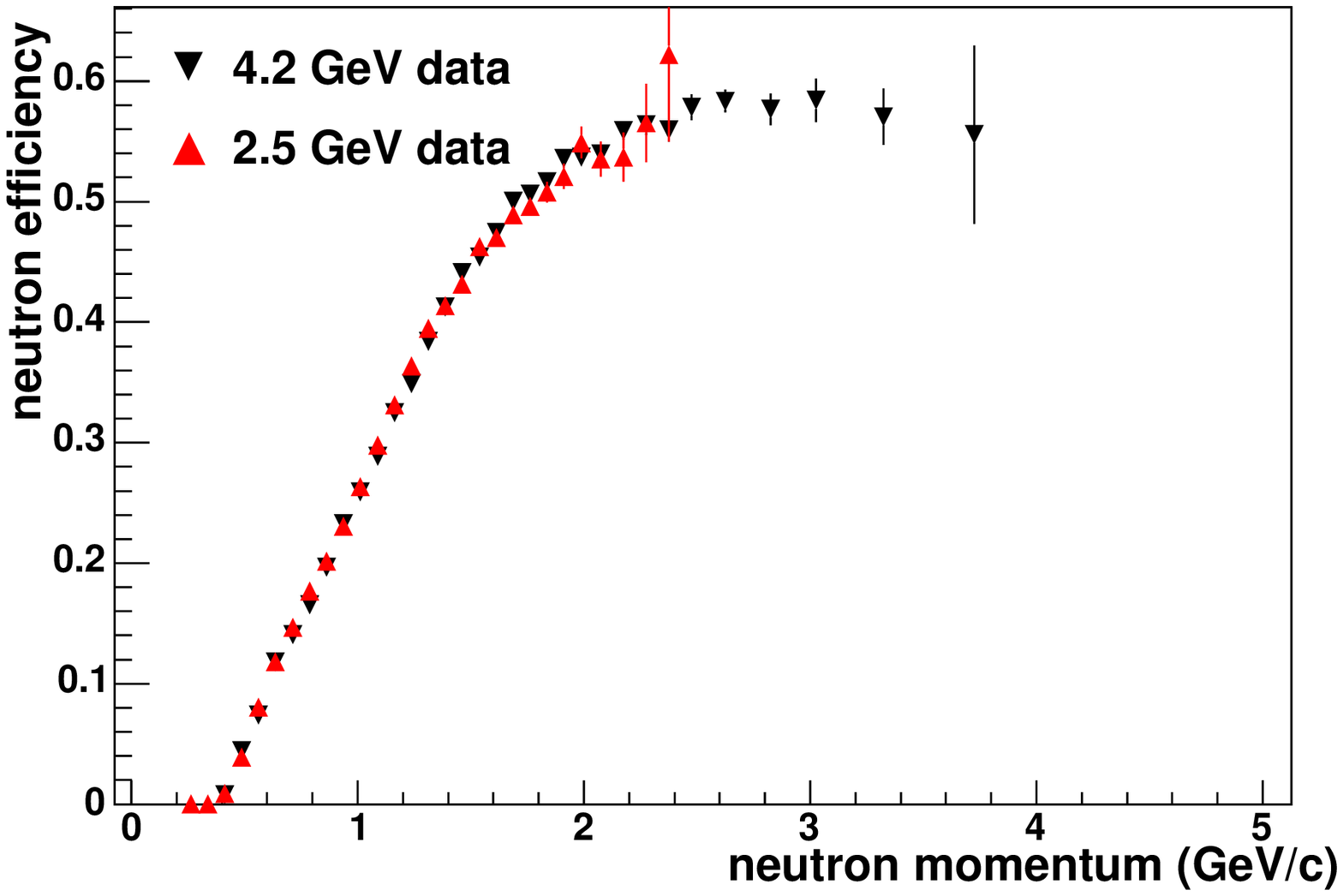} \put(0.0,0){\includegraphics[height=1.8in]{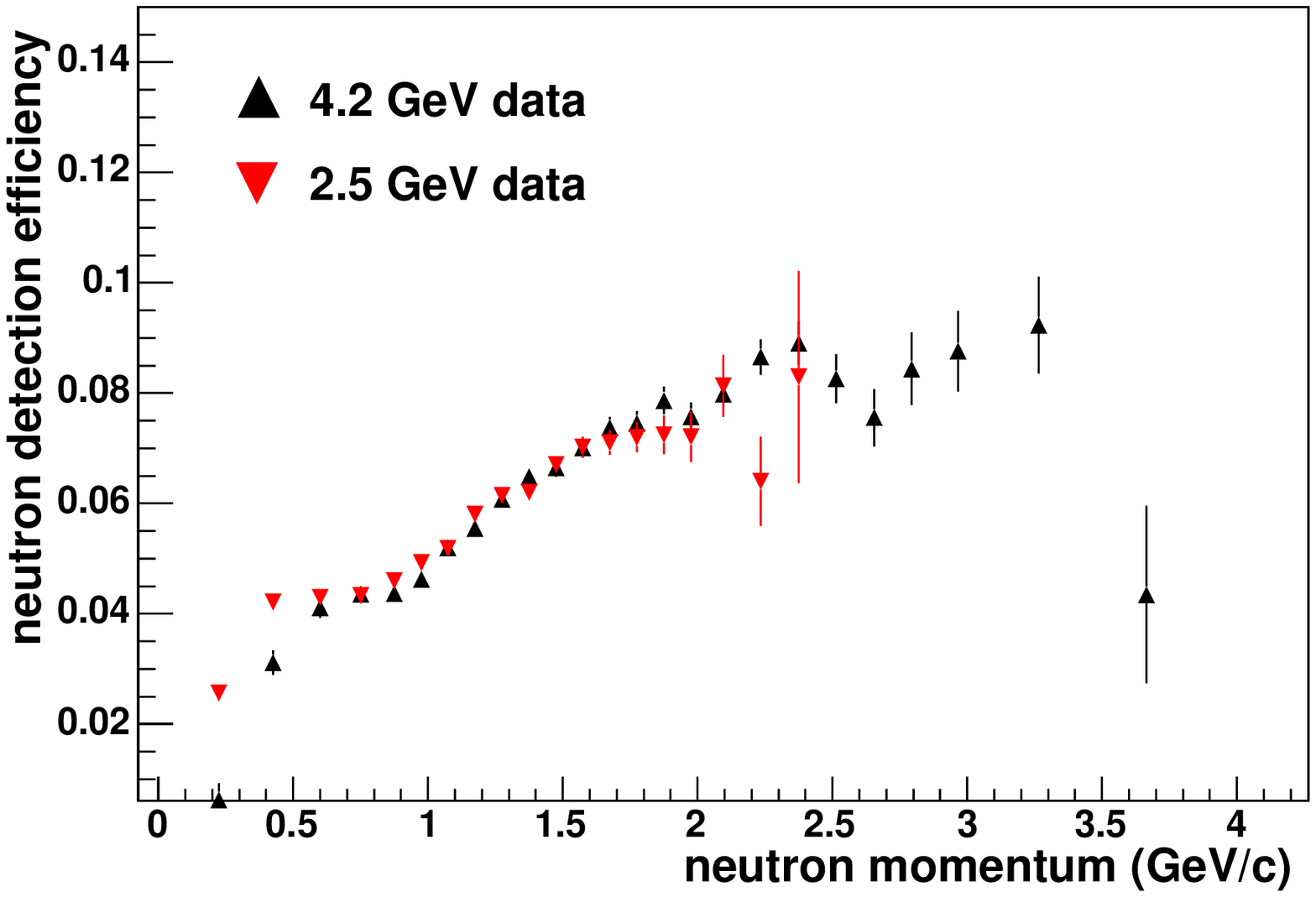}}
\caption{Detection efficiency versus momentum for 
neutrons detected in the forward-angle electromagnetic calorimeters at two 
different beam energies (left-hand panel) and in the TOF system (right-hand panel). 
The efficiency has been integrated 
over all six sectors of the CLAS.}\label{fig:brooks_nde}
\end{figure}

Neutrons were measured in two CLAS scintillator-based detectors: the
forward-angle electromagnetic shower calorimeters and
the time-of-flight (TOF) scintillators. The efficiency measurement was
performed using tagged neutrons from the $^1\rm{H}(e,e'\pi^+)X$ reaction
where the mass of the final state $M_X$ was chosen to be that of the
neutron. 
Since the precise value of the detection efficiency can vary with
time-dependent and rate-dependent quantities such as 
photomultiplier tube gain, the detection efficiency was measured
\emph{simultaneously} with the primary deuterium measurement. Two separate
targets were positioned in the beam at the same time, one for
deuterium and the other for hydrogen, separated by less than 5 cm.
Plots of the resulting neutron detection efficiencies are shown in
Fig. \ref{fig:brooks_nde}. The left-hand  plot shows the results for the forward
electromagnetic shower calorimeter, while the right-hand panel shows the results
for the time of flight scintillators.

\begin{figure}[htb]
\begin{center}
\includegraphics[scale=0.46,angle=0]{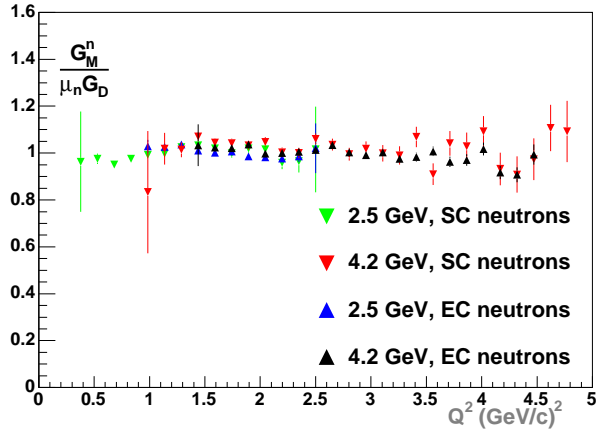}
\caption{\label{fig:gmn_detector_compare}Preliminary results for 
$G_M^n / (\mu_n G_{D}) $ as a function of $Q^2$ for four 
different measurements Uncertainties are statistical one only.}\label{fig:brooks_results}
\end{center}
\end{figure}

The CLAS extraction of $G_M^n(Q^2)$ actually consists of multiple
overlapping measurements. The time of flight scintillators cover the
full angular range of the spectrometer, while the forward calorimeters cover
a subset of these angles, thus $G_M^n(Q^2)$ can be obtained from two
independent measures of the neutron detection efficiency. In addition,
the experiment was carried out with two different beam energies that
had overlapping coverage in $Q^2$, so that the detection of the
protons of a given $Q^2$ took place in two different regions of the
drift chambers. As a result, essentially four measurements of
$G_M^n(Q^2)$ have been obtained from the CLAS data that
could have four independent sets of systematic errors. 
Preliminary results are shown in Figure \ref{fig:brooks_results} of
the reduced form factor $G_M^n /(\mu_n G_D)$ for the measurements.
The four measurements are consistent within the statistical errors,
suggesting that the systematic errors are well-controlled and small.

One of the goals of this experiment is to achieve a systematic uncertainty of 3\% or less \cite{E94-017}.
The biggest contributor to this uncertainty is the parameterization of the neutron detection
efficiency, but there are significant contributions from the uncertainty in the other elastic form factors
(recall Equation 1), and the effect of
the Fermi motion in the deuteron.
Details on the determination of these uncertainties and other smaller ones can be found in Ref \cite{jl1}.
Here we discuss the analysis of the large contributors.

As described above the neutron detection efficiency in the calorimeters and the TOF system was measured 
simultaneously with the production data using tagged neutrons from the $p(e,e^\prime \pi^+)n$
reaction.
The results for both detector systems were fitted with a polynomial at low neutron momentum 
and a plateau at large
momentum. The order of the polynomial and position of the plateau edge were varied
to test the sensitivity of $G_M^n$ as a function of $\rm Q^2$.
Uncertainties in the range 1-2\% were obtained.
The uncertainties in the other elastic form factors contribute to the
uncertainty in $G_M^n$ (see Equation 1).
The uncertainty in the proton cross section was estimated using the difference
between two parameterizations by Bosted and Arrington \cite{bosted, arrington}.
For the effect of $G_E^n$, the difference between the Galster parameterization
and a fit by Lomon was used \cite{galster,lomon}.
These uncertainties had a maximum of 1.5\% and were typically much less.
The other large contributor was the effect of the Fermi motion in the
deuteron knocking the scattered nucleons out of the CLAS acceptance.
The effect was studied in a simulation using two dramatically different
choices for the Fermi momentum distribution; a flat distribution and the 
Hulthen distribution.
These two Fermi momentum distributions have very different effects on the
neutron and proton spectra, but in the ratio we found the difference to be 
less than 1\%.
The complete inventory of uncertainties was combined in a weighted average of the
systematic uncertainty as a function of $\rm Q^2$.
The final uncertainty varied from 1.7-2.5\% across the full $\rm Q^2$ range of the data.

\begin{figure}[htb]
\begin{center}
\includegraphics[scale=0.5,angle=0]{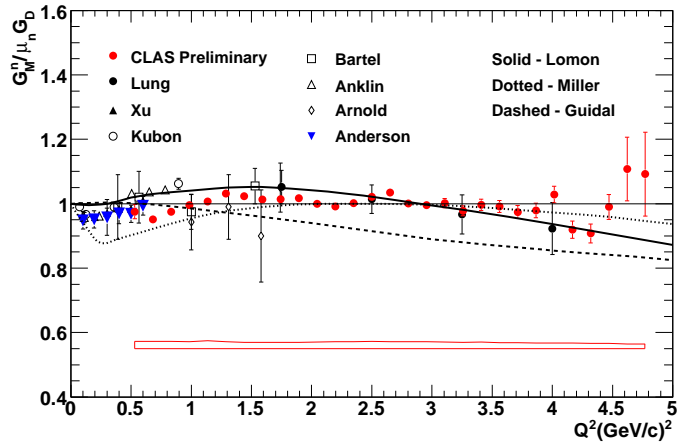}
\caption{Preliminary results for $G_M^n / (\mu_n G_{D}) $ from CLAS
are compared with a selection of previous data. See Ref \cite{Kees} and references
therein and Ref. \cite{anderson}}\label{fig:final_results}
\end{center}
\end{figure}

The preliminary results are shown in Fig. \ref{fig:final_results}
together with a sample of existing data. The
error bars shown on the points are due only to statistical uncertainties. The bar graph represents
the systematic uncertainty as a function of $\rm Q^2$. The data
shown are the weighted averages of the four overlapping individual
measurements 
discussed above. 
A few features are noteworthy. 
First, the quality and coverage of the
data is a very substantial improvement over the existing world's data
set. Second, the dipole form appears to give a good
representation of the data over the $Q^2$ range measured, which is at
variance at higher $Q^2$ with parameterizations based on
previous data, which tend to show a more strongly
decreasing trend for $G_M^n /(\mu_n G_{D})$ with increasing $Q^2$.


The curves shown in Figure \ref{fig:final_results} are from
theoretical calculations by Lomon, Guidal, {\it et al.}, and Miller \cite{lomon,guidal,miller1}
using the world's data on the elastic form factors without the experimental
results we report here.
In the Lomon model the $\rho$, $\omega$, $\phi$, $\omega^\prime (1419)$, and $\rho^\prime(1450)$
vector meson pole contributions evolve at high momentum
transfer to conform to the predictions of perturbative QCD.
Recent data on the elastic nucleon form factors measured with polarization techniques
is included in the sample.
Excellent fits are obtained when older data on $G_E^n$ and $G_E^p$ inconsistent with the 
recent polarization results are excluded.
In the work by Guidal, {\it et al.} a Regge parameterization of the generalized
parton distribution (GPDs) is used to characterize the elastic nucleon form factors at
low momentum transfer and then extended to higher momentum transfer. 
The calculation reproduces the more rapid drop observed in existing data at higher $\rm Q^2$,
but is not consistent with the dipole approximation and our preliminary results.
In the Miller work, the nucleon is treated using light-front dynamics as a relativistic
system of three bound quarks and a surrounding pion cloud.
The model achieves a good description of the existing nucleon form factors, but does
not include the results here in the analysis.

\section{Virtual Photon Asymmetry of the Proton}

The spin structure of the nucleon has been investigated in
a series of much-discussed polarized lepton scattering experiments
over the last 25 years \cite{dharma, moskov, zheng, fatemi, yun}. These measurements, most
of which covered the deep inelastic scattering (DIS) region of
large final-state invariant mass W and momentum transfer $\rm Q^2$,
compared the $\rm Q^2$-dependence of the polarized structure function
$g_1$ with perturbative QCD evolution equations and shed new light on
the structure of the nucleon. Among the most surprising results
was the realization that only a small fraction of the nucleon spin
(20-30\%) is carried by the quark helicities, in disagreement
with quark model expectations of 60-75\%. This reduction is
often attributed to the effect of a negatively polarized quark sea
at low momentum fraction x, which is typically not included in
quark models (see the paper by Isgur \cite{nathan} for a more detailed discussion).

For a more complete understanding of the quark structure of
the nucleon, it is advantageous to concentrate on a kinematic region
where the scattering is most likely to occur from a valence
quark in the nucleon carrying more than a fraction $x = 1/3$ of
the nucleon momentum. In particular, the virtual photon asymmetry,
$A_1(x) \approx g_1(x)/F_1(x)$ (where $F_1$ is the usual unpolarized
structure function) can be (approximately) interpreted in
terms of the polarization $\Delta u/u$ and $\Delta d/d$ of the valence $u$ and
$d$ quarks in the proton in this kinematic region, while the contribution
from sea quarks is minimized. This asymmetry also
has the advantage of showing only weak $\rm Q^2$-dependence [6,8],
making a comparison with various theoretical models and predictions
more straightforward.

In this Proceedings, we discuss the first high-precision measurement
of $A_1(x,Q^2)$ for the proton and the deuteron at moderate to
large $x$ ($x\ge 0.15$) over a range of momentum transfers 
$\rm Q^2 = 0.05$-$\rm 5.0 ~ GeV^2$, covering both the resonance and the deep
inelastic region \cite{dharma}.
Longitudinally polarized electrons from CEBAF of several beam energies
around 1.6 GeV and 5.7 GeV were scattered off longitudinally
polarized ammonia targets—$\rm ^{15}NH_3$ and $\rm ^{15}ND_3$ and detected
in CLAS.
The target material was kept in a $1~ K$ liquid helium bath and
was polarized via Dynamic Nuclear Polarization (DNP) \cite{tgt}.
The target polarization was monitored online using a Nuclear
Magnetic Resonance (NMR) system. The beam polarization
was measured at regular intervals with a Moeller polarimeter.
The product of beam and target polarization ($P_bP_t$ ) was determined
from the well-known asymmetry for elastic (quasielastic)
scattering from polarized protons (deuterons), measured
simultaneously with inelastic scattering. For the 1.6 GeV data
set, the average polarization product was $P_b P_t = 0.540\pm 0.005$
($0.180 \pm 0.007$) for the $\rm ^{15}NH_3$ ($\rm ^{15}ND_3$) target. The corresponding
values for the 5.7 GeV data set are $0.51 \pm 0.01$ and
$0.19\pm 0.02$.

The data analysis proceeds along the following steps (see
Refs. \cite{dharma, yun} for more details). We first extract the raw count rate asymmetry
$A^{raw}_{||} = (N^+ − N^−)/(N^+ + N^−)$, where the electron
count rates for anti-parallel ($N^+$) and parallel ($N^−$) electron
and target polarization are normalized to the (live-time gated)
beam charge for each helicity. The background due to misidentified
pions and electrons from decays into $e^+e^−$ pairs (a few
percent in all cases) has been subtracted from these rates. We
divide the result by the product of beam and target polarization
$P_bP_t$ and correct for the contribution from non-hydrogen
nuclei in the target. For this purpose, we use auxiliary measurements
on $\rm ^{12}C$, $\rm ^4He$ and pure $\rm ^{15}N$ targets. We then combine the
asymmetries for different beam and target polarization directions,
thereby reducing any systematic errors from false asymmetries
(no significant differences between the different polarization
sets were found). Finally we apply radiative corrections
using the code RCSLACPOL \cite{radcor}. The (quasi-)elastic
radiative tail contribution to the denominator of the asymmetry
is treated as a further dilution factor.

The final result is the longitudinal (Born) asymmetry 
$A_{||} = D(A_1 + \eta A_2)$, where the depolarization factor 
$D = (1 - E^\prime \epsilon/E)/(1 + \epsilon R)$, $E$ is the beam energy, $E^\prime$ is the scattered electron
energy, $\epsilon = (2EE^\prime − Q^2/2)/(E^2 + {E^\prime}^2 + Q^2/2)$ is the virtual
photon polarization, $R < 0.2$ is the ratio of the longitudinal
to the transverse photoabsorption cross section and
$\eta = (\epsilon\sqrt{Q^2} )/(E − E^\prime\epsilon)$. 
The asymmetry $A_2$ is the longitudinal-transverse interference
virtual photon asymmetry.
We use the standard notations
for the energy transfer, $\nu = E − E^\prime$, and four-momentum
transfer squared, $Q^2 = 4EE^\prime \sin^2(\theta/2)$.
A parameterization of the world's data was used to model $A_2$ and $R$ and to extract
$A_1$ \cite{radcor, par1}.

\begin{figure}[htb]
\begin{center}
\includegraphics[scale=0.7,angle=0]{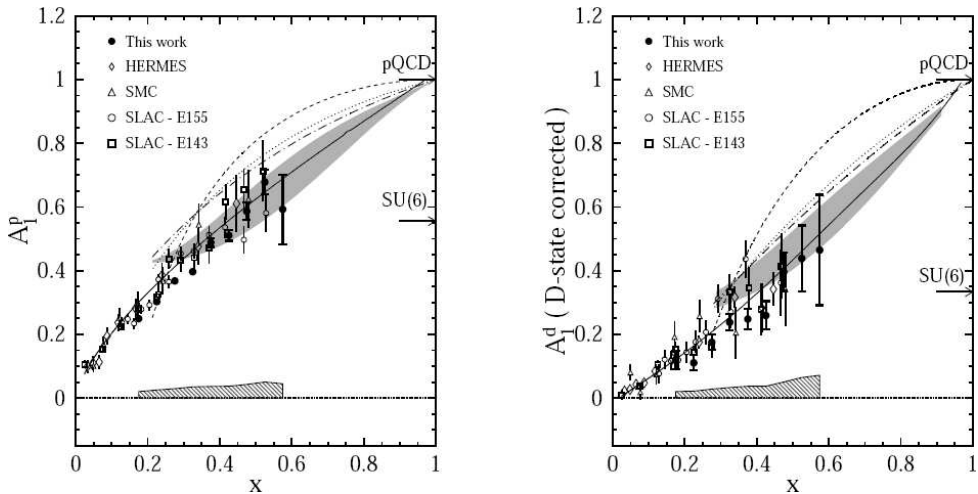}
\caption{Results for the asymmetry $A_1(x)$ for protons (left-hand panle) and deuterons 
(right-hand panel) \cite{dharma}. The $SU(6)$ expectation is shown by the arrow in each panel.}\label{fig:A1pd}
\end{center}
\end{figure}

The results for $A_1(x)$, averaged over $\rm Q^2 > 1~ GeV^2$ and
$W >2 ~ \rm{GeV}$, are shown in Fig. \ref{fig:A1pd} for the proton and  
the deuteron. 
At small $x$, where our average $\rm Q^2$ is close to
$\rm 1~ GeV^2$, the data fall below our parametrization of the world
data with $\rm Q^2 = 10~ GeV^2$ (solid line). This deviation is due
to the $\rm Q^2$-dependence shown in Ref\cite{dharma}. In contrast,
all data points for the proton and the deuteron lie above the
SU(6) values for $x >0.45$. The hyperfine interaction model of
SU(6) symmetry breaking by Isgur \cite{nathan} (grey band in figures) is
closest to the data. Of the different mechanisms for SU(6) symmetry
breaking considered by Close and Melnitchouk \cite{cm}, the
model with suppression of the symmetric quark wave function
(dot-dashed curve in Fig \ref{fig:A1pd}) deviates least from the data. 
The dashed curve is for a model using helicity-1/2 dominance and the dotted one
is for spin-1/2 dominance.
In general, our results are in better agreement with models (like
the first two mentioned above) in which the ratio of down to up
quarks, $d/u$, goes to zero and the polarization of down quarks,
$\Delta d/d$ tends to stay negative for rather large values of $x$, in contrast
to the behavior expected from hadron helicity conservation.

\begin{figure}[htb]
\begin{center}
\includegraphics[height=3.5in]{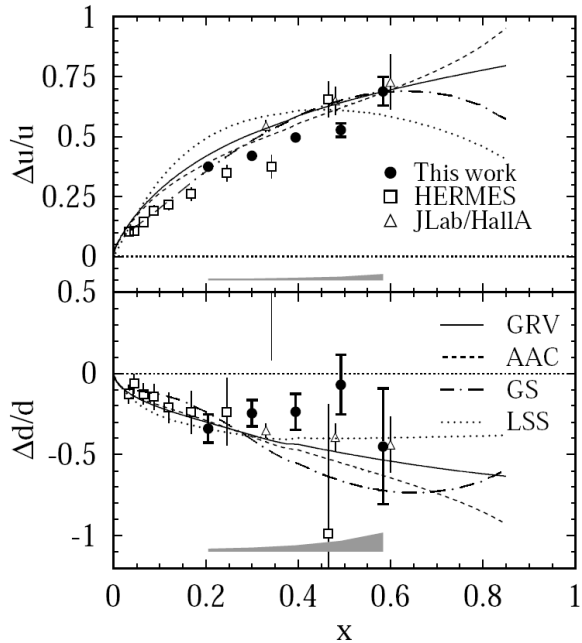}
\caption{Quark polarizations $\Delta u /u$ (upper panel) and $\Delta d/d$ (lower panel)
extracted from the asymmetry data \cite{dharma}.}\label{fig:qdf1}
\end{center}
\end{figure}

Within a naive quark - parton model (and ignoring any contribution
from strange quarks), we can estimate the quark (plus
antiquark) polarizations $\Delta u/u$ and $\Delta d/d$ directly from our
data by combining the results for $g_1$ from the proton and the
deuteron (including some nuclear corrections for the deuteron
D-state and Fermi motion) with our parametrization of the
world data on $F_1^p$ and $F_1^d$.
The result shown in Figure \ref{fig:qdf1} has relatively large statistical errors for
$\Delta d/d$, since neither $A_1^p$ nor $A_1^d$
are very sensitive to $\Delta d/d$.
Included are all data
above $W = 1.77 ~\rm{GeV}$ and $\rm Q^2 = 1~ GeV^2$. Also shown are semi-inclusive results
from HERMES \cite{hermes} and inclusive results from Hall A data [11] combined with
previous data from CLAS \cite{fatemi}. The solid line is from the leading order fit to the world
data by GRSV \cite{grsv}, the dashed line is from the AAC fit \cite{aac}, the dash-dotted
line is from Gehrmann and Stirling \cite{gehrmann} and the dotted line indicates the latest
fit from LSS \cite{lss} which includes higher twist corrections.
Our estimate is
consistent with the result from the $\rm ^3He$ experiment \cite{zheng}, showing
no indication of a sign change to positive values up to
$x \approx 0.6$. At the same time, these data for $\Delta u/u$  show a consistent
increase with x, compatible with $\Delta u/u \rightarrow 1$ as $x\rightarrow 1$. Our data
are also in reasonable agreement with existing leading order perturbative QCD fits.


\section{The Future}

The future of the study of form factors and structure functions at JLab.
As part of the 6-GeV program at JLab analysis continues on a low-$\rm Q^2$ measurement of 
the magnetic form factor of the neutron and in other areas like the
transition form factors which probe the structure of the excited states of the nucleon. New exit channels
are under investigation to extract the structure functions.
More importantly, both the study of structure functions and form factors will be an essential component
of the 12-GeV Upgrade at JLab.
This new project will double the current energy of CEBAF, add a new 
experimental hall to focus on the discovery of exotic mesons, and upgrade the other three, existing
halls to take advantage of the new physics opportunities.
CLAS will be replaced by a new detector CLAS12 which will be able to operate at a luminosity ten times
greater than the current device.
The scientific motivation for measuring the magnetic form factor of the neutron and the virtual photon
asymmetry is strongly made in the Conceptual Design Report \cite{cdr}.
In fact, two experiments to measure $A_1(x)$ (PR12-06-109) and $G_M^n$ (PR12-07-104) have been approved in the last eighteen 
months by the JLab Program Advisory Committee for running during the first five years of the 
Upgrade \cite{pac30, pac32}.

\end{document}